\documentstyle[aps,multicol,epsf,amsbsy]{revtex}
\newcommand{\kB}{k_{\mathrm{B}}}

\begin{document} 
\title{Magnetic relaxation in a classical spin chain as model for
  nanowires} 
\author{D.~Hinzke and U.~Nowak}
\address{Theoretische Tieftemperaturphysik,
  Gerhard-Mercator-Universit\"{a}t-Duisburg, 47048 Duisburg/ Germany\\ 
  e-mail: denise@thp.uni-duisburg.de, uli@thp.uni-duisburg.de }
\date{August 9, 1999}
\maketitle
\begin{abstract}
  With decreasing particle size, different mechanisms dominate the
  thermally activated magnetization reversal in ferromagnetic
  particles.  We investigate some of these mechanisms for the case of
  elongated, single-domain nanoparticles which we describe by a
  classical Heisenberg spin chain driven by an external magnetic
  field. For sufficiently small system size the magnetic moments
  rotate coherently.  With increasing size a crossover to a reversal
  due to soliton-antisoliton nucleation sets in. For even larger
  systems many of these soliton-antisoliton pairs nucleate at the same
  time. These effects give rise to a complex size dependence of the
  energy barriers and characteristic time scales of the relaxation.  We
  study these quantities using Monte Carlo simulations as well as a
  direct integration of the Landau-Lifshitz-Gilbert equation of motion
  with Langevin dynamics and we compare our results with asymptotic
  solutions for the escape rate following from the Fokker-Planck
  equation. Also, we investigate the crossover from coherent rotation
  to soliton-antisoliton nucleation and multi-droplet nucleation,
  especially its dependence on the system size, the external field and
  the anisotropy of the system.
\end{abstract}

Pacs: 75.10.Hk, 75.40.Mg, 75.40.Gb

\begin{multicols}{2}
\section{Introduction}
Many novel physical effects occur in connection with the decreasing
size of the systems which are under investigation. Magnetic materials
are now controllable in the nanometer regime and there is a broad
interest in the understanding of the magnetism of small magnetic
structures and particles due to the broad variety of industrial
applications \cite{himpsel}.  Magnetic particles which are small
enough to be single-domain are proposed to have good qualities for
magnetic recording and arrays of isolated, nanometer-sized particles
are thought to enhance the density of magnetic storage. But on the
other hand there is an ultimate limit for the density of magnetic
storage which is given by that size of the particles below which 
superparamagnetism sets in \cite{chantrell}. Therefore, the role of
thermal activation for the stability of the magnetization in
nanometer-sized structures and particles is studied at present
experimentally as well as theoretically.

Wernsdorfer et al. measured the switching times in isolated
nanometer-sized particles \cite{wernsdorfer_pa,wernsdorfer_nu}, and
wires \cite{wernsdorfer_wi_prl,wernsdorfer_wi_prb}. For small enough
particles \cite{wernsdorfer_pa} they found agreement with the theory
of N\'{e}el \cite{neel} and Brown \cite{brown} who described the magnetization
switching in Stoner-Wohlfarth particles by thermal activation over a
single energy barrier following from coherent rotation of the magnetic
moments of the particle. For larger particles \cite{wernsdorfer_nu}
and also for wires \cite{wernsdorfer_wi_prl,wernsdorfer_wi_prb}
nucleation processes and domain wall motion were found to become
relevant.

Asymptotic formulae for the escape rates following from corresponding
Fokker-Planck equations have been derived for ensembles of isolated
Stoner-Wohlfarth particles
\cite{neel,brown,klik,braun_cr,coffey,garanin} as well as for a
one-dimensional model \cite{braun,braun_prb,braun_cr,braun_oe}.

Most of the numerical studies of the magnetization switching base on
Monte Carlo methods. Here, mainly nucleation phenomena in Ising models
have been studied
\cite{stauffer,rikvold,rikvold_coer,acharyya,hinzke,nowak} but also
vector spin models have been used
\cite{gonzales_prb,gonzales_japp,hinzke,nowak} to investigate the
magnetization reversal in systems with many, continuous degrees of
freedom qualitatively. However, Monte Carlo methods - even though well
established in the context of equilibrium thermodynamics - do not
allow for a quantitative interpretation of the results in terms of a
realistic dynamics. Only recently, a Monte Carlo method with a
quantified time step was introduced \cite{nowak_algo}. Here, the
interpretation of a Monte Carlo step as a realistic time interval was
achieved by a comparison of one step of the Monte Carlo process with a
time interval of a Langevin equation, i.e., a stochastic
Landau-Lifshitz--Gilbert equation.

Numerical methods for the direct integration of a Langevin equation
\cite{lyberatos,garcia} are more time consuming than a Monte
Carlo method but nevertheless highly desirable since here, naturally,
a realistic time is introduced by the equation of motion.  The
validity of different integration schemes is still under discussion
(see \cite{garcia} for a discussion of the validity of It\^{o} and
Stratonovich integration schemes) and, hence, here, as well as for the
Monte Carlo methods, the investigation of analytically solvable models
as test tools for the evaluation of the numerical techniques are
desirable.

In this paper we will consider a chain of classical magnetic moments
--- a system which can be interpreted as a simplified model for
ferromagnetic nanowires or extremely elongated nanoparticles
\cite{braun}. This model is very useful for two reasons: i) since it
was treated analytically asymptotic results for the energy barriers as
well as for the escape rates are available. Hence, the model can be
used as a test tool for numerical techniques. ii) Depending on the
system size, given anisotropies and the strength of the magnetic field
different reversal mechanisms may appear and can be investigated. For
small system sizes the magnetic moments are expected to rotate
coherently \cite{neel}, while for sufficiently large system sizes the
so-called soliton-antisoliton nucleation is proposed \cite{braun}.
Therefore, a systematic numerical investigation of the relevant energy
barriers, time scales, and of the crossover from coherent rotation to
nucleation is possible.

The outline of the paper is as follows. In Sec.~\ref{s:sim} first we
introduce the model and we compare the two different numerical
techniques mentioned above, namely the Monte Carlo method and the
numerical solution of the Langevin equation, both of which we will use
throughout the paper. In Sec.~\ref{s:theo} we compare our numerical
results with the theoretical considerations of Braun, concerning the
energy barriers as well as the mean first passage times for
magnetization switching by coherent rotation as well as by
soliton-antisoliton nucleation. For higher temperatures or driving
fields, we also find a crossover to multidroplet nucleation, similar
to what is known from Ising models. In section \ref{s:con} we
summarize our results and relate them to experimental work.

\section{Model and Methods}
\label{s:sim}
\subsection{Model}                         
\label{s:model}
Our intention is to compare our numerical results with Braun's
analytical work which bases on a continuum model for a magnetic
nanowire. For our numerical investigations we use a discretized
version of this model namely a one dimensional classical Heisenberg
model. We consider a chain of magnetic moments of length $L$ (number
of spins) with periodical boundary conditions defined by the
Hamiltonian,
\begin{eqnarray}
  \label{e:ham}
  E &=& - J \sum_{\langle ij \rangle} {\mathbf S}_i \cdot {\mathbf S}_j
  -d_x \sum_i (S^x_i)^2  \nonumber \\
  &+& d_z \sum_i (S^z_i)^2- \mu_s{\mathbf B} \cdot \sum_i {\mathbf S}_i. 
\end{eqnarray}
where the ${\mathbf S}_i = {\boldsymbol \mu}_i/\mu_s$ are three
dimensional magnetic moments of unit length. The first sum which
represents the exchange of the magnetic moments is over nearest
neighbor interactions with the exchange coupling constant $J$.  The
second and third sum represent uniaxial anisotropies of the system
where the $x$-axis is the easy axis and the $z$-axis the hard axis of
the system (anisotropy constants $d_x = 0.1J$, $d_z = J$). These
anisotropy terms may contain contributions from shape anisotropy as
well as crystalline anisotropies \cite{braun_oe}.  Even though an
exact treatment of dipolar interactions would be desirable we let this
problem for future work so that our results presented here are
comparable to the analytical work of Braun.  The last sum is the
coupling of the moments to an external magnetic field, where ${\mathbf
  B}$ is the induction. 

All our simulations using the two different methods above start with a
configuration where all magnetic moments point into the $x$ direction,
antiparallel to the external magnetic field ${\mathbf B} = -B_x
{\mathbf \hat{x}}$.

The time $\tau$ when the $x$-component of the magnetization changes
its sign averaged over many simulation runs (100-1000, depending on
system size and simulation method) is the most important quantity
which we determine.  In the case where the temperatures is low
compared to the energy barrier the system is in the metastable initial
state for a very long time while the time needed for the magnetization
reversal itself is extremely short. In this case $\tau$ can be
considered to be the so-called mean first passage time required to
overcome the energy barrier. For low enough temperatures the mean
first passage time should also be comparable to the reciprocal of the
escape rate $\lambda$ which has been calculated for the model
considered here asymptotically from the Fokker-Planck equation in
certain limits (for details see \cite{braun_prb}).  In the classical
nucleation theory the so-called lifetime or nucleation time is the
time required by the system to build a supercritical droplet which
from then on will grow systematically and reverse the system. The
metastable lifetime was measured numerically in Ising models by use of
similar methods \cite{stauffer,rikvold,rikvold_coer,acharyya,hinzke,nowak}.
For low enough temperature $T$ all the different times mentioned above
are expected to coincide.

\subsection{Langevin dynamics}                   
The equation describing the dynamics of a system of magnetic moments
is the Landau-Lifshitz-Gilbert (LLG) equation of motion with Langevin
dynamics. This equation has the form
\begin{eqnarray}
    \label{e:ll}
  {\textstyle \frac{(1+\alpha^2)\mu_s}{ \gamma}}\frac{\partial {\bf
  S}_i}{\partial t} &=& {\mathbf S}_i \times \Big ({\boldsymbol \zeta}_{i}(t) -
  \frac{\partial E}{\partial {\mathbf S}_i}\Big) \nonumber \\
  &-& \alpha {\bf S}_i \times \Big({\mathbf S}_i\times \big({\boldsymbol \zeta}_{i}(t)
    - \frac{\partial E}{\partial {\mathbf S}_i}\big)\Big),
\end{eqnarray}
with the gyromagnetic ratio $\gamma = 1.76 \times 10^{11}
(\mbox{Ts})^{-1} $, the dimensionless damping constant $\alpha$.
The first part of Eq.~\ref{e:ll} describes the spin precession while
the second part includes the relaxation of the moments. In both parts
of this equation the thermal noise ${\boldsymbol \zeta}_i(t)$ is
included representing thermal fluctuations, with $ \langle {\boldsymbol
  \zeta}_i(t) \rangle = 0$ and $ \langle {\boldsymbol \zeta}_i^k(t)
{\boldsymbol \zeta}_j^l(t') \rangle = 2 \delta_{ij} \delta_{kl}
\delta(t-t')2 \alpha \kB T \mu_{s}/ \gamma$ where $i, j$ denote the
lattice sites and $k, l$ the cartesian components.
Eq.~\ref{e:ll} is solved numerically using the Heun
method which corresponds to
the Stratonovich discretization scheme \cite{garcia}.  Note, that
since the Langevin dynamics simulations are much more time consuming
than the Monte Carlo simulation we use this method here mainly for
comparison and we will present only data for the relaxation on shorter
time scales.

\subsection{Monte Carlo simulations}            
\label{s:mc}
It is inconceivable that the Langevin dynamics simulation can be used
over the whole time scale of physical interest so that we simulate the
system also using Monte Carlo methods \cite{binder} with a heat-bath
algorithm.  Monte Carlo approaches in general have no physical time
associated with each step, so that an unquantified dynamic behavior is
represented. However, we use a new time quantified Monte Carlo method
which was proposed in \cite{nowak_algo} where the interpretation of a
Monte Carlo step as a realistic time interval was achieved by a
comparison of one step of the Monte Carlo process with a time interval
of the LLG equation in the high damping limit.  We will use this
algorithm in the following.  The trial step of this algorithm is a
random movement of a magnetic moment within a cone of size $r$ with
\begin{equation}
  r^2 = \frac{20 k_B T \alpha \gamma}{(1+\alpha^2) \mu_s} \Delta t.
  \label{e:trial}
\end{equation}
Using this algorithm one Monte Carlo step represents a given time
interval $\Delta t$ of the LLG equation in the high damping limit as
long as $\Delta t$ is chosen appropriately (for details see
\cite{nowak_algo}).

\narrowtext
\begin{figure}[h]
  \begin{center}
    \epsfysize=4.5cm
    \epsffile{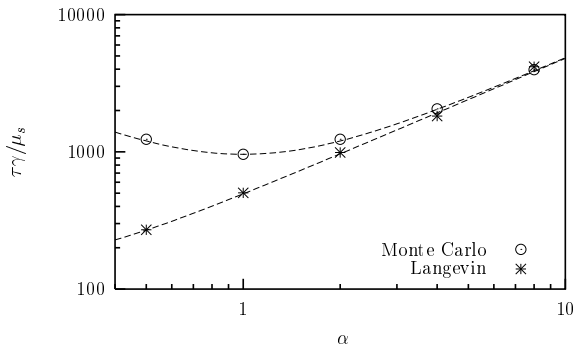}
  \end{center}
  \caption{Reduced mean first passage time $\tau\gamma/\mu_s$
    vs. damping constant $\alpha$. The data are from Monte Carlo and
    Langevin dynamics simulations for $\kB T = 0.025 J$ and $L = 80$.
    Solid lines are guides to the eye.}
 \label{f:alpha}
\end{figure}

To test the algorithm, in Fig.~\ref{f:alpha} the $\alpha$ dependence
of the mean first passage time of the Monte Carlo data is compared
with the data of the Langevin dynamics simulation.  Each data point is
an average over 1000 independent runs. For low values of the damping
constant the data do not coincide while in the high damping limit
Monte Carlo and Langevin dynamics data converge.  Hence, throughout
this work we will use $\alpha = 4$, which is large enough so that our
Monte Carlo simulation and the numerical solution of Eq.~\ref{e:ll}
yield identical time scales. Even though this is an unphysically large
value for $\alpha$ we can compare our results with the analytically
obtained high damping asymptotes.

\section{Results}
\label{s:theo}
We investigate the influence of the system size on the occurring
reversal mechanisms. There are two extreme cases of reversal
mechanisms which might occur in our model namely coherent (or uniform)
rotation and nucleation. For small system sizes all magnetic moments
of the particle rotate uniformly in order to minimize the exchange
energy. For larger system sizes it is favorable for the system to
divide into parts of opposite directions of magnetization parallel to
the easy axis which minimizes the anisotropy energy.  This is a
magnetization reversal driven by nucleation and subsequent domain wall
propagation. The crossover between these mechanisms is discussed later
in this section.  First, we study the different reversal mechanism and
compare our numerical data with theoretical formulae.

\subsection{Coherent Rotation}         
\label{s:cr}
In the case of small chain length the magnetic moments rotate
coherently while overcoming the energy barrier which is due to the
anisotropy of the system. The first theoretical description of the
coherent rotation of elongated single-domain particles was developed
by Stoner and Wohlfarth \cite{stoner}. The belonging energy barrier is
\begin{equation}
  \Delta E_{\mathrm cr} = Ld_x(1-h)^2
  \label{e:e_cr}
\end{equation}
with $h = \mu_s B_x/(2d_x)$.  N\'{e}el expanded this model for the
case of thermal activation \cite{neel} and Brown \cite{brown}
calculated the escape time
\begin{equation}
  \tau = \tau^*_{\mathrm cr} \exp{\frac{\Delta E_{\mathrm cr}}{\kB T}}
  \label{e:act}
\end{equation}
following a thermal activation law. The prefactor $\tau^*$ is not a
simple constant attempt frequency as claimed frequently by several
authors but a complicated function which in general may depend on
system size, temperature, field and anisotropies. For our model it is
\begin{eqnarray}
 \frac{\tau^*_{\mathrm cr} \gamma}{\mu_s} = {\textstyle \frac{\pi 
    (1+\alpha^2)\sqrt{\frac{d_z(1+h)}{d_x(1-h)+d_z}}}{\alpha
    (d_x(1-h^2)-d_{z}) +
    \sqrt{(\alpha (d_x(1-h^2)+ d_z))^2+4d_x d_z(1-h^2)}}} 
  \label{e:pre_cr}
\end{eqnarray}
as was calculated from the Fokker-Planck equation \cite{braun_cr}.

The main assumption underlying all results above is that the system can
be described by a single degree of freedom, namely the magnetic moment
of the particle which must have a constant absolute value.  Then, the
equations above should hold for low enough temperatures $\kB T \ll
\Delta E_{\mathrm cr}$ and for $\mu_s B_x < 2d_x $.  For larger fields
the energy barrier is zero, so that the reversal is spontaneous
without thermal activation (non-thermal magnetization reversal).

\begin{figure}[h]
  \begin{center}
    \epsfysize=4.5cm
    \epsffile{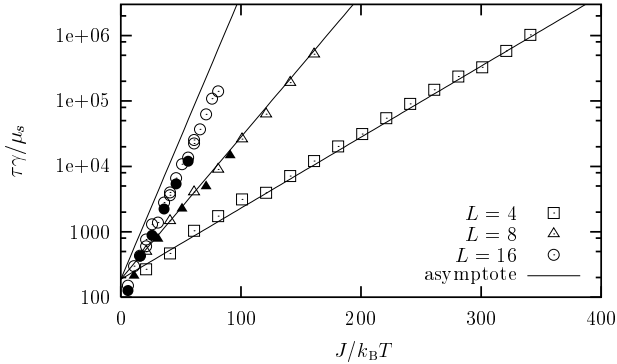}
  \end{center}
  \caption{Reduced mean first passage time $\tau\gamma/\mu_s$ vs. inverse
    temperature. The data are from Monte Carlo (open symbols) and
    Langevin dynamics (filled symbols) simulations for different
    system sizes. The solid lines are the asymptotic formulae for the
    escape times (see Eq.~\ref{e:act} and Eq.~\ref{e:pre_cr}).}
    \label{f:tau_cr}
\end{figure}

In the following, our numerical results for the mean first passage
times $\tau$ for coherent rotation are compared with the equations
above.  Fig.~\ref{f:tau_cr} shows the temperature dependence of $\tau$
for a given value of the external magnetic field $h=0.75$ and three
different system sizes.  For temperatures $\kB T < \Delta E_{\mathrm
  cr}$ our data confirm the asymptotic solutions above for smaller
system sizes.

\narrowtext

For the largest system shown here ($L = 16$) the numerical data are
systematically lower than the theoretical prediction. Obviously, the
prefactor $\tau^*_{\mathrm cr}$ depends on the system size in
contradiction to Eq.~\ref{e:pre_cr} while the energy barrier
(Eq.~\ref{e:e_cr}) is obviously still correct (the slope of the
data). This size dependence of $\tau^*_{\mathrm cr}$ can be explained
with the temperature dependence of the absolute value of the
magnetic moment of the extended system. In an extended system, the
energy barrier for vanishing field, $d_x L$, is reduced to $\textstyle
d_x \langle \sum_i (S^x_i)^2 \rangle$ which expanded to first order of
the temperature can lead to corrections of the form $\textstyle d_x
\langle \sum_i (S^x_i)^2 \rangle \approx d_x L(1-aT)$.  Including this
in Eq.~\ref{e:act} remarkably leads to an effective reduction of the
prefactor $\tau^*_{\mathrm cr}$ by a factor $\exp(-a d_x L)$ and not
to a reduction of the effective energy barrier $\Delta E_{\mathrm
  cr}$. To conclude, in extended systems with many degrees of freedom
the reduction of the magnetization leads effectively to a size
dependent reduction of the prefactor of the thermal activation law for
low temperatures.

\subsection{Soliton-Antisoliton Nucleation}          
\label{s:SAN}
With increasing length of the chain a different reversal mechanism
becomes energetically favorable, namely the so-called
soliton-antisoliton nucleation proposed by Braun \cite{braun_prb}.
Here, during the reversal the system splits into two parts with
opposite directions of magnetization parallel to the easy axis. These
two parts are separated by two domain walls with opposite directions
of rotation in the easy $x-y$-plane (a soliton-antisoliton pair) which
pass the system during the reversal. 

Note, that we consider periodic boundary conditions,
otherwise also the nucleation of one single soliton at one end of
the chain would be possible.  The energy barrier
$\Delta E_{\mathrm nu}$ which has to be overcome during this
nucleation process is \cite{braun_prb}
\begin{equation}
  \Delta E_{\mathrm nu} = \sqrt{2Jd_{x}}\big(4\mbox{tanh}R - 4hR\big),
  \label{e:e_nu}
\end{equation}
with $R = \mbox{arcosh}(\sqrt{1/h})$. In the limit of small magnetic
fields, $h \to 0$, this energy barrier has the form $ \Delta E_{\mathrm
  nu} = 4\sqrt{2Jd_{x}}$ which represents the well-known energy of two
domain walls.  The corresponding mean first passage time follows once
again a thermal activation law (see Eq.~\ref{e:act}).  Since our Monte
Carlo simulations are for rather large damping, $\alpha = 4$, we
compare our numerical data with the prefactor obtained in the
over-damped limit (Eq.~5.4 in \cite{braun_prb}) which in our units is
\begin{eqnarray}
  \frac{\tau^*_{\mathrm nu} \gamma}{\mu_s} = \frac{\pi^{3/2}
    (1+\alpha^2) (\kB T)^{1/2} (2J)^{1/4}} {16 \alpha \; L \;
    d_x^{7/4} |E_0(R)| \tanh R^{3/2} \sinh R}
  \label{e:pre_nu}
\end{eqnarray}
The eigenvalue $E_0(R)$ has been calculated numerically in
\cite{braun_prb}. In the limit $h \to 1$ it is $|E_0(R)|\approx 3R^2$.
The $1/L$ dependence of the prefactor reflects the size dependence of
the probability of nucleation. The larger the system the more probable
is the nucleation process. Furthermore, the prefactor has a remarkable
$\sqrt{T}$ dependence leading to a slight curvature in the
semi-logarithmic plot of the thermal activation law (see
Eq.~\ref{e:act}). 

\narrowtext
\begin{figure}[h]
  \begin{center}
    \epsfysize=5cm
    \epsffile{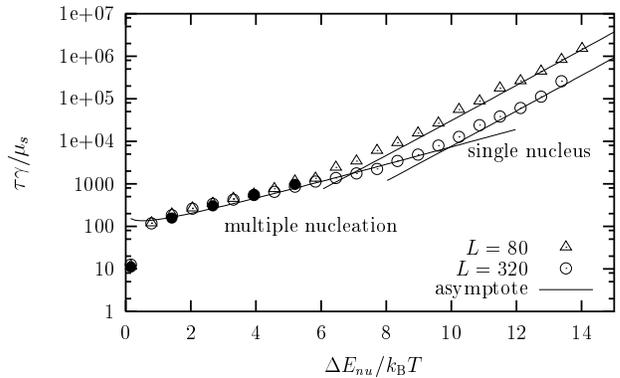}
  \end{center}
  \caption{Reduced mean first passage time $\tau\gamma/\mu_s$
    vs. inverse temperature $\Delta E_{\mathrm nu}/\kB T$ for two
    different system sizes. The data are from Monte Carlo (open
    symbols) and Langevin dynamics (filled symbols) simulations. The
    solid lines are the asymptotic formulae for the inverse escape
    rates for a single soliton-antisoliton nucleation process (see
    Eqs.~\ref{e:act}, \ref{e:e_nu}, and \ref{e:pre_nu}) and multiple
    nucleation (see Eq.~\ref{e:tau_md}).}
    \label{f:tau_nu}
\end{figure}

Fig.~\ref{f:tau_nu} shows the temperature dependence of the reduced
mean first passage time for $h = 0.75$ and two different system sizes.
The formulae above are confirmed only for rather low temperatures
($\Delta E_{\mathrm nu}/\kB T > 8$ for the smaller system and $\Delta
E_{\mathrm nu}/\kB T > 10$ for the larger system).  In the range of
intermediate temperatures (but still $\kB T < \Delta E_{\mathrm nu}$)
the numerical data of both, Langevin dynamics and Monte Carlo
simulations deviate from the formulae above.  Interestingly, in this
region the mean first passage times do not depend on system size in
contrast to Eq.~\ref{e:pre_nu} where a $1/L$-dependence occurs due to
the size dependence of the nucleation probability.

\narrowtext
\begin{figure}  
  \begin{center}
    \epsfysize=2.1cm
    \epsffile{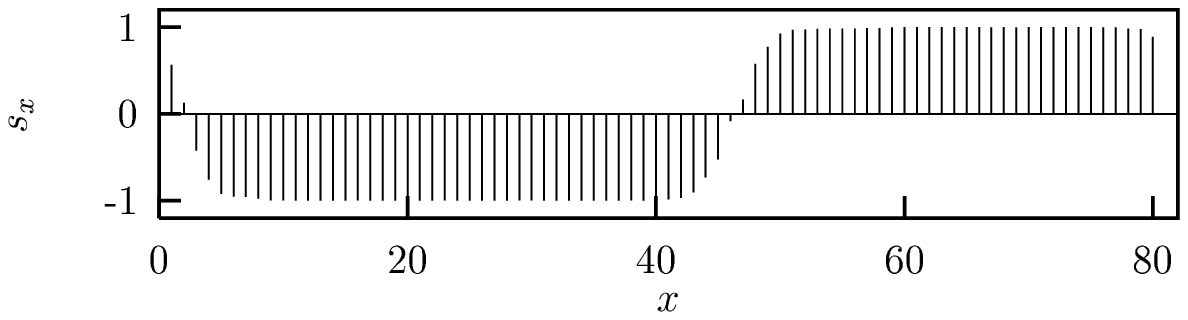}
    \vspace{0.5cm}
        \epsfysize=2.1cm
     \epsffile{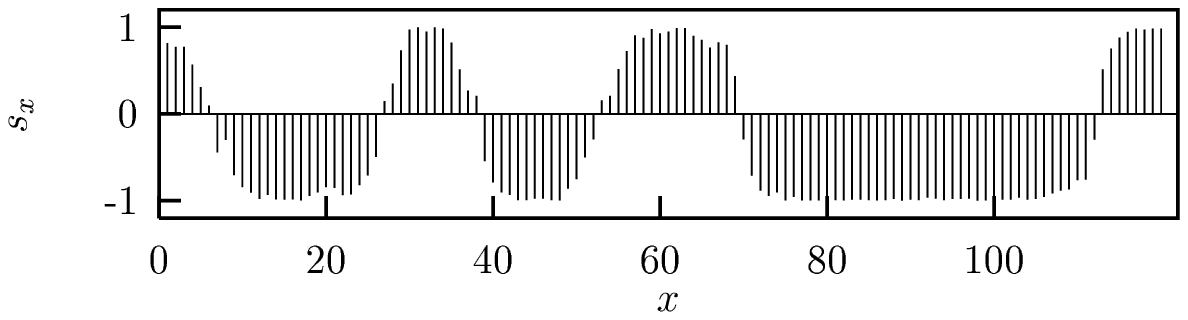}
  \end{center}
  \caption{$x$-component of the magnetic moments of the chain at the
    time $\tau$ after single soliton-antisoliton nucleation (upper
    diagram, $L = 80$, $\kB T = 0.0019J$, $h = 0.95$) and multiple
    nucleation ($L=120$, $\kB T = 0.038J$, $h = 0.95$)}
  \label{f:pic}
\end{figure}

In order to understand this effect occurring in the intermediate
temperature range, Fig.~\ref{f:pic} shows the $x$-component of the
magnetic moments of the chain at the mean first passage time $\tau$.
The upper diagram shows the pure soliton-antisoliton nucleation
occurring at low temperatures.  Due to thermal fluctuations a nucleus
is originated and two domain walls pass the system. The lower diagram
shows that alternatively also several nuclei may grow simultaneously.
Obviously, depending on temperature (and also other quantities, like
system size and field) with a certain probability many nuclei may
arise at the same time.  This is a multiple nucleation which was
investigated mainly in the context of Ising models \cite{rikvold_arc}
where it is called {\em multidroplet nucleation}.

\subsection{Multidroplet Nucleation}                  
\label{s:MDN}
The mean first passage time $\tau_{\mathrm mn}$ for the multidroplet
nucleation can be calculated with the aid of the classical nucleation
theory \cite{becker,rikvold_arc}.  In the case of multiple nucleation
many nuclei with a critical size originate within the same time.
These supercritical nuclei grow and join each other leading to the
magnetization reversal. The mean first passage time for this process
is determined by the probability for the occurrence of supercritical
nuclei $1/\tau_{\mathrm nu}$ and the time that is needed for the
growth of the nuclei.  Let us assume that the radius of a
supercritical nucleus grows linearly with the time $t$. Then
$\tau_{\mathrm mn}$ is given by the condition that the change of the
magnetization $\Delta M$ equals the system size \cite{acharyya},
\begin{equation}
  \Delta M(\tau_{\mathrm mn}) = \int_{0}^{\tau_{\mathrm
      mn}} \frac{(2vt)^D}{\tau_{\mathrm nu}} \mbox{d}t = L^D,
\end{equation}
where $v$ is the domain wall velocity and $D$ the dimension of the
system.  Hence, the time when half of the system is reversed is given by
\begin{equation}
  \tau_{\mathrm mn} = \big(\frac{L}{2v}\big)^{\frac{D}{D+1}} \big(
  (D+1)\tau^*_{\mathrm nu} \big)^{\frac{1}{D+1}}\exp{\frac{\Delta
      E_{\mathrm nu}}{(D+1)\kB T}}.
\end{equation}
For the one dimensional system  which we consider in this paper 
the lifetime is given by
\begin{equation}
  \tau_{\mathrm mn} = \sqrt{\frac{L \tau^*_{\mathrm nu}}{v}} \exp{\frac{\Delta
      E_{\mathrm nu}}{2\kB T}}.
\label{e:tau_md}
\end{equation}
This means that the (effective) energy barrier for the multidroplet
nucleation is reduced by a factor 1/2 and the prefactor $\tau_{\mathrm
  mn}^*$ does no longer depend of the system size since
$\tau^*_{\mathrm nu}$ for the soliton-antisoliton nucleation has a
$1/L$-dependence (see Eq.~\ref{e:pre_nu}).  Since we do not know the
value of the domain wall velocity we had to fit this parameter which
influences the prefactor $\tau^*_{\mathrm mn}$ for the comparison in
Fig.~\ref{f:tau_cr} in the intermediate temperature range.
Nevertheless, in Fig.~\ref{f:tau_cr} the reduction of the slope of the
curve by the factor 1/2 in the multidroplet regime is confirmed. Also,
it can be seen, that the prefactor is not dependent on the system
size since the data for both sizes coincide as is explained by the
considerations above (we used the same value for $v$ of course for
both curves, since $v$ should not depend on the system size).

A similar crossover from single to multidroplet excitations was
observed in Ising models, field dependent
\cite{rikvold,rikvold_coer,acharyya} as well as temperature dependent
\cite{nowak}. Comparing the mean first passage times for single and
multiple nucleation we get for the intersection of these two times the
crossover condition
\begin{eqnarray}
  L_{\mathrm sm} = \sqrt{v\tau^*_{\mathrm nu} L_{\mathrm sm}
    }\exp{\frac{\Delta E_{\mathrm nu}}{2 \kB T}}.
  \label{e:cross_sm}
\end{eqnarray}
The corresponding time is $L_{\mathrm sm}/v$ --- the time that a
domain wall needs to cross the system. This results is also comparable
to calculations in Ising models \cite{rikvold}.

\subsection{Crossover}                     
In this section we are interested in the crossover between coherent
rotation and soliton-antisoliton nucleation. Therefore, we compare the
energy barrier of soliton-antisoliton nucleation (see
Eq.~\ref{e:e_nu}) with the energy barrier of coherent rotation (see
Eq.~\ref{e:e_cr}) in order to obtain a condition for the mechanism
with the lowest activation energy. The resulting crossover line $L_c$
has the form

\begin{eqnarray}
  L_c = \frac{\sqrt{2Jd_{x}}\big(4\mbox{tanh}R - 4hR\big)}{d_x(1-h)^2}.
  \label{e:crossover}
\end{eqnarray}
For vanishing magnetic field it is $L_c = 4 \sqrt{\frac{2J}{d_x}}$ and
we get the simple condition that four domain walls (not two!) have to
fit into the system. Note, that Braun also determined this crossover
line \cite{braun_cross} from a slightly different condition. He
calculated the system size below which no nonuniform solution for the
Euler-Lagrange equations exists and he obtained a limit which is
similar to Eq.~\ref{e:crossover} but a little bit lower.

\begin{figure}[h]
  \begin{center}
    \epsfxsize=9cm
    \epsffile{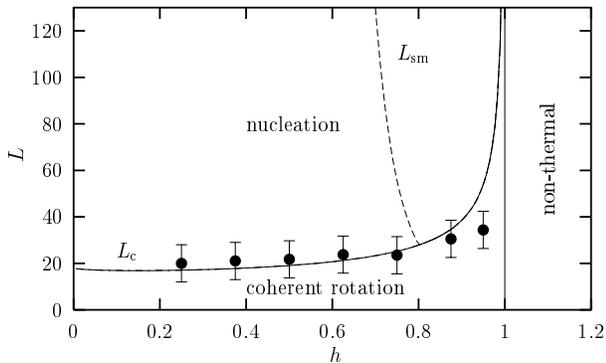}
  \end{center}
  \caption{The diagram shows the regions of different reversal
    mechanisms. The lines corresponds to Eq.~\ref{e:crossover} and
    \ref{e:cross_sm}. The points are data from Monte Carlo simulations
    confirming Eq.~\ref{e:crossover}.}
 \label{f:phase}
\end{figure}

A diagram showing which reversal mechanisms occur in our model
depending on system size and field is presented in Fig.~\ref{f:phase}.
The crossover line $L_c$ derived above separates the coherent rotation
region from that of soliton-antisoliton nucleation.  For $h > 1$ the
reversal is non-thermal. In the nucleation region, for larger
fields a temperature dependent crossover to multiple nucleation sets
in.  However, the lower the temperature the more vanishes this region.
As an example we show in the figure the line $L_{\mathrm sm}$ which
separates single from multiple nucleation for $\kB T = 0.006J$ and under
the assumptions that the domain wall velocity is proportional to the
field \cite{rikvold}.

\narrowtext

\begin{figure}  
  \begin{center}
    \epsfysize=4.5cm
    \epsffile{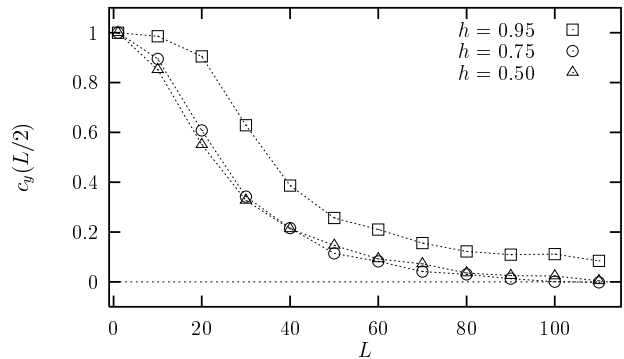}
  \end{center}
  \caption{System size dependence of $c_y(L/2)$ for different magnetic
    fields. $\kB T$ = 0.1 \dots 0.004$J$ depending on $h$.}
  \label{f:ord}
\end{figure}

In order to confirm Eq.~\ref{e:crossover} numerically we determine the
mean correlations of the $y$-components of two spins located in a
distance of $L/2$ (the maximum distance in a system with periodic
boundary conditions) at the time $\tau$ during the reversal
\begin{eqnarray*}
  c_y(L/2) = \Big[ \frac{1}{N} \sum_{i=1}^N
  S_i^y(\tau)S_{i+L/2}^y(\tau) \Big]_{\mathrm av}.
\end{eqnarray*}
This quantity characterizes the reversal mechanism.  Fig.~\ref{f:ord}
shows the system size dependence of $c_y(L/2)$ for different values of
the magnetic field. For coherent rotation the limiting value of
$c_y(L/2)$ is 1 since at the time $\tau$ all moments point into the
$y-$direction (small systems).  On the other hand, for nucleation the
system is mainly split into two parts where the moments point into
$\pm x-$ direction. In this case the correlations in $y-$ direction
are zero (larger system size). As a criterion for the crossover
between nucleation and coherent rotation we use a value of $1/2$ for
the correlation function and define the corresponding system size as
the crossover length. This analysis leads to the numerical data shown
in Fig.~\ref{f:phase}.

\narrowtext

The MC data confirm the theoretical crossover line except for the
largest magnetic field. Here, the numerical data differ from the
theoretical crossover line which diverges in the limit $h \to 1$.  In
this region a mixed reversal mechanism appears: first the magnetic
moments rotate coherently up to a certain rotation angle and then an
instability sets in and a (restricted) soliton-antisoliton pair
arises. Obviously, the coherent rotation is unstable in this region. A
hint for the existence of this mixed mechanism can also be found in
Fig.~\ref{f:ord}: for the field $h = 0.95$ the correlation
is not zero for large system sizes but remains finite due to the fact
that the whole system first rotates by a certain angle before the
nucleation sets in.

\section{Conclusions}
\label{s:con}
We investigated the magnetization switching in a classical Heisenberg
spin chain which we consider as simple model for nanowires or
elongated nanoparticles and as a test tool for numerical techniques
since analytical expressions for the relevant energy barriers and time
scales exist in several limits. Numerically we used Monte Carlo
methods as well as Langevin dynamics simulations and we confirmed that
the Monte Carlo algorithm we use yield time quantified data comparable
to those of the Langevin dynamics simulation in the limit of high
damping.

Varying the system size we observed different reversal mechanisms,
calculated the mean first passage times for the reversal (i.e. the
time scale of the relaxation process) and compared our results with
analytical considerations.  The comparison of our numerical data with
theoretical considerations confirms that both of our numerical
techniques can be used to study the relaxation behavior of magnetic
systems qualitatively and quantitatively.

\narrowtext
\begin{figure}  
  \begin{center}
    \epsfxsize=8.cm
    \epsffile{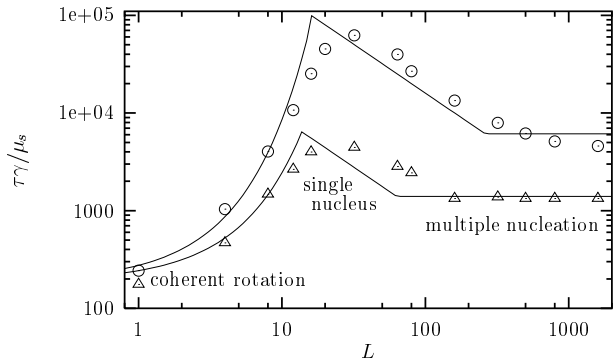}
  \end{center}
  \caption{Reduced mean first passage time $\tau\gamma/\mu_s$
    vs. system size $L$ for two different temperatures ($\kB T = 0.024
    J$ and $\kB T = 0.016J$). The solid lines are the appropriate
    formulae and the data are from Monte Carlo simulations.}
  \label{f:tau_L}
\end{figure}

Fig.~\ref{f:tau_L} summarizes our results. It shows the system size
dependence of the reduced mean first passage time for two different
temperatures.  For small system sizes the spins rotate coherently.
Here, the energy barrier (Eq.~\ref{e:e_cr}) is proportional to the
system size leading to an increase of $\tau$ with system size.
Following Eq.~\ref{e:pre_cr} the prefactor of the thermal activation
law should not be $L$ dependent.  However, here we found slight
deviations from the asymptotic expressions stemming from the
non-constant magnetization of extended systems.  In the regime of
soliton-antisoliton nucleation the energy barrier (Eq.~\ref{e:e_nu})
does not depend on the system size but the prefactor
(Eq.~\ref{e:pre_nu}) has a $1/L$ dependency.  Interestingly, this
leads to a decrease of the mean first passage time with increasing
system size. Therefore, there is a maximum relaxation time close to
where the crossover from coherent rotation to nucleation occurs (see
Eq.~\ref{e:crossover}). This decrease ends where the so-called
multidroplet nucleation sets in (Eq.~\ref{e:cross_sm}). From here on
with increasing system size the mean first passage time remains
constant (Eq.~\ref{e:tau_md}).

Note, that qualitatively the same behavior can be found in the
particle size dependence of the dynamic coercivity. Solving the
equation describing the thermal activation (Eq.~\ref{e:act}) in the
three regimes explained above for $h(L, \tau = \mbox{const})$, one
finds an increase of the coercivity in the coherent rotation regime, a
decrease in the nucleation regime, and at the end a constant value for
multiple nucleation. This is qualitatively in agreement with
measurements of the size dependence of barium ferrite recording
particles \cite{chang}.

Comparing the energy barrier for the nucleation process with
experimental values (Eq.~\ref{e:e_nu}), one should first mention that
in an experimental system nucleation usually will start a the sample
ends. This will reduce the energy barrier by a factor of 2 since
without periodical boundary conditions one does not need a
soliton-antisoliton pair but just one single excitation.  The energy
barrier for the nucleation process (Eq.~\ref{e:e_nu}) was compared in
\cite{braun_oe} with energy barriers measured in isolated Ni-nanowires
\cite{wernsdorfer_wi_prl,wernsdorfer_wi_prb} finding the experimental
energy barrier reduced by a factor of 3. This agreement is rather
encouraging taking into account that in realistic systems the energy
barrier might be reduced depending on the form of the sample ends.

Regarding the crossover from coherent rotation to nucleation a similar
analysis, even though less rigorous, has been performed for three
dimensional systems \cite{hinzke}. Here, in the limit $h\to0$ the
crossover diameter is $L_c = 3/2 \; \sqrt{2J/d_x}$ instead of $2 \;
\sqrt{2J/d_x}$ for the spin chain with open boundary conditions. As
discussed in \cite{hinzke} for CoPt particles the crossover length
should be of the order of 30nm which is also roughly in agreement with
experiments on Co particles \cite{wernsdorfer_pa,wernsdorfer_nu}.

We should mention that the model treated here contains magnetostatic
energy only in a local approximation \cite{braun_oe}. Arguments
against this treatment have been brought forward \cite{aharoni} so
that the question wether an explicit inclusion of the dipole-dipole
interaction would lead to additional reversal modes (e.~g. curling) -
also depending on wether the diameter of the nanowire is larger or
smaller than the exchange length - remains open.  Calculations
following these lines are therefore highly desirable.

\section*{Acknowledgments}
We thank K.~D.~Usadel and D.~A.~Garanin for fruitful discussion and
H.~B.~Braun also for providing numerical results for the eigenvalue
$|E_0(R)|$. This work was supported by the Deutsche
Forschungsgemeinschaft through the Graduiertenkolleg "Struktur und
Dynamik heterogener Systeme" and was done within the framework of the COST
action P3 working group 4.

\bibliographystyle{local/aip}
\bibliography{Cite}

\begin{thebibliography}{10}

\bibitem{himpsel}
F.~J. Himpsel, J.~E. Ortega, G.~J. Mankey, and R.~F. Willis,
\newblock Adv.~Phys {\bf 47}, 511 (1998).

\bibitem{chantrell}
R.~W. Chantrell and K.~O'Grady,
\newblock The magnetic properties of fine particles,
\newblock in {\em Applied Magnetism}, edited by R.~Gerber, C.~D. Wright, and
  G.~Asti, page 113, Kluwer Academic Publishers, Dordrecht, 1994.

\bibitem{wernsdorfer_pa}
W.~Wernsdorfer et~al.,
\newblock Phys.~Rev.~Lett. {\bf 78}, 1791 (1997).

\bibitem{wernsdorfer_nu}
W.~Wernsdorfer et~al.,
\newblock Phys.~Rev.~B {\bf 53}, 3341 (1996).

\bibitem{wernsdorfer_wi_prl}
W.~Wernsdorfer et~al.,
\newblock Phys.~Rev.~Lett. {\bf 77}, 1873 (1996).

\bibitem{wernsdorfer_wi_prb}
W.~Wernsdorfer et~al.,
\newblock Phys.~Rev.~B {\bf 55}, 11552 (1997).

\bibitem{neel}
L.~N\'{e}el,
\newblock Ann.~Geophys. {\bf 5}, 99 (1949).

\bibitem{brown}
W.~F. Brown,
\newblock Phys.~Rev {\bf 130}, 1677 (1963).

\bibitem{klik}
I.~Klik and L.~Gunther,
\newblock J.~Stat.~Phys. {\bf 60}, 473 (1990).

\bibitem{braun_cr}
H.~B. Braun,
\newblock J.~Appl.~Phys {\bf 76}, 6310 (1994).

\bibitem{coffey}
W.~T. Coffey et~al.,
\newblock Phys.~Rev.~Lett. {\bf 80}, 5655 (1998).

\bibitem{garanin}
D.~A. Garanin, E.~C. Kennedy, D.~S.~F. Crothers, and W.~T. Coffey,
\newblock cond-mat/9903192  (1999).

\bibitem{braun}
H.~B. Braun,
\newblock Phys.~Rev.~Lett. {\bf 71}, 3557 (1993).

\bibitem{braun_prb}
H.~B. Braun,
\newblock Phys.~Rev.~B {\bf 50}, 16485 (1994).

\bibitem{braun_oe}
H.~B. Braun,
\newblock J.~Appl.~Phys {\bf 85}, 6127 (1999).

\bibitem{stauffer}
D.~Stauffer,
\newblock Int.~J.~Mod.~Phys. C {\bf 3}, 1059 (1992).

\bibitem{rikvold}
P.~A. Rikvold, H.~Tomita, S.~Miyashita, and S.~W. Sides,
\newblock Phys.~Rev.~E {\bf 49}, 5080 (1994).

\bibitem{rikvold_coer}
P.~A. Rikvold, M.~A. Novotny, M.~Koleski, and H.~L. Richards,
\newblock Nucleation theory of magnetization switching in nanoscaled
  ferromagnets,
\newblock in {\em Dynamical properties of unconventional magnetic systems},
  edited by A.~T. Skjeltrop and D.~Sherrington, page 307, Kluwer, Dordrecht,
  1998.

\bibitem{acharyya}
M.~Acharyya and D.~Stauffer,
\newblock Euro.~Phys.~J.~B {\bf 5}, 571 (1998).

\bibitem{hinzke}
D.~Hinzke and U.~Nowak,
\newblock Phys.~Rev.~B {\bf 58}, 265 (1998).

\bibitem{nowak}
U.~Nowak and D.~Hinzke,
\newblock J.~Appl.~Phys {\bf 85}, 4337 (1999).

\bibitem{gonzales_prb}
J.~M. Gonz\'{a}les, R.~Ram\'{\i}rez, R.~Smirnov-Rueda, and J.~Gonz\'{a}lez,
\newblock Phys.~Rev.~B {\bf 52}, 16034 (1995).

\bibitem{gonzales_japp}
J.~M. Gonz\'{a}les, R.~Smirnov-Rueda, and J.~Gonz\'{a}lez,
\newblock J.~Appl.~Phys {\bf 81}, 5573 (1997).

\bibitem{nowak_algo}
U.~Nowak, R.~W. Chantrell, and E.~C. Kennedy,
\newblock cond-mat/9906089  (1999).

\bibitem{lyberatos}
A.~Lyberatos and R.~W. Chantrell,
\newblock J.~Appl.~Phys {\bf 73}, 6501 (1993).

\bibitem{garcia}
J.~L. Garc\'{i}a-Palacios and F.~J. L\'{a}zaro,
\newblock Phys.~Rev.~B {\bf 58}, 14937 (1998).

\bibitem{binder}
K.~Binder and D.~W. Heermann,
\newblock in {\em Monte Carlo Simulation in Statistical Physics}, edited by
  P.~Fulde, page~21, Springer-Verlag, Berlin, 1997.

\bibitem{stoner}
E.~C. Stoner and E.~P. Wohlfarth,
\newblock Philos.\ Trans.\ R.\ Soc.\ A {\bf 240}, 599 (1949).

\bibitem{rikvold_arc}
P.~A. Rikvold and B.~M. Gorman,
\newblock Recent results on the decay of metastable phases,
\newblock in {\em Annual Reviews of Computational Physics I}, edited by
  D.~Stauffer, page 149, World Scientific, Singapore, 1994.

\bibitem{becker}
R.~Becker and W.~D{\"{o}}ring,
\newblock Ann.~Physik {\bf 24}, 719 (1935).

\bibitem{braun_cross}
H.~B. Braun,
\newblock World Scientific, Singapore, 1999.

\bibitem{chang}
T.~Chang, J.-G. Zhu, and J.~H. Judy,
\newblock J.~Appl.~Phys {\bf 73}, 6716 (1993).

\bibitem{aharoni}
A.~Aharoni,
\newblock J.~Appl.~Phys {\bf 80}, 3133 (1996).

\end{thebibliography}

\end{multicols}

\end{document}